# Scalable fabrication of edge contacts to 2D materials

*Naveen Shetty[1], Hans He,[1,2] Richa Mitra[1], Johanna Huhtasaari[1], Konstantina Iordanidou[3], Julia Wiktor[3], Sergey Kubatkin[1], Saroj Dash[1], Rositsa Yakimova[4], Lunjie Zeng[3], Eva Olsson[3], and Samuel Lara-Avila[1,5,]\**

[1] Department of Microtechnology and Nanoscience, Chalmers University of Technology, 412 96 Gothenburg, Sweden

[2] RISE Research Institutes of Sweden, Box 857, S-50115 Borås, Sweden

[3] Department of Physics, Chalmers University of Technology, 412 96 Gothenburg, Sweden

[4]Department of Physics, Chemistry and Biology, Linköping University, 581 83 Linköping, Sweden

[5]National Physical Laboratory, Hampton Road, Teddington TW11 0LW, United Kingdom

*Author to whom correspondence should be addressed: samuel.lara@chalmers.se

KEYWORDS. Edge-contacts, 2Dmaterial, MoS$_2$, graphene, epitaxial graphene

We report a reliable and scalable fabrication method for producing electrical contacts to two-dimensional (2D) materials, based on the tri-layer resist system. We demonstrate the applicability of this method for epitaxial graphene on silicon carbide (epigraphene) and the transition metal dichalcogenides (TMDCs) molybdenum disulfide ($MoS_2$). For epigraphene, data on nearly 70 contacts result in median values of the specific contact resistances $\rho_c \sim 67$ $\Omega\mu m$, and follow the Landauer quantum limit, $\rho_c \sim n^{-1/2}$, with $n$ being the carrier density of graphene. For $MoS_2$ flakes, our edge contacts enable field effect transistors (FET) with ON/OFF ratio of $> 10^6$ at room temperature ($> 10^9$ at cryogenic temperatures). The fabrication route here demonstrated allows for contact metallization using thermal evaporation and also by sputtering, giving an additional flexibility when designing electrical interfaces, which is key in practical devices and when exploring the electrical properties of emerging materials.

Two dimensional (2D) materials stand as one of the of most interesting technological platforms for the development of novel and disruptive electronics[1–3]. Being atomically thin, the electron transport properties of 2D materials are extremely responsive to externally applied stimuli (e.g. light, electric field, magnetic field), making them ideal candidates for implementation of transducers with unprecedented sensitivity. Additionally, the high mobility demonstrated in some 2D materials, notably graphene, allow for devices operating at very high frequencies that are not attainable in a straightforward manner with bulk materials. Yet, a notable and often overlooked development required for the wide spread use of future electronics based on 2D materials is to improve the quality, reproducibility and reliability of the electrical contacts[4–7]. In practice, electrical interfaces to any electronic device should satisfy the following requirements: a) allow supplying the necessary current to a device, b) the voltage drop across the contacts should be small compared to the voltage drop across the active device region, and c) should be stable over time and in different environments. To this day, there are some examples of successful strategies to form electrical contacts to 2D materials. However, the choice of contact materials for a particular 2D material remains somewhat of an art[8–10], and a truly scalable fabrication of reproducible contacts to a wide variety of 2D materials remains to be demonstrated[11].

First proposed for carbon nanotubes, edge-contacts (a.k.a. end-contacts) [12–15] is currently one of the most successful and popular methods to achieve good electrical interfacing to 2D materials. In the most common implementation of this method, originally demonstrated by the Columbia group for graphene [15], the 2D layer is encapsulated by hexagonal boron nitride (hBN). The entire stack is then patterned and etched to expose only the edge of the graphene layer, which is in subsequently metalized, to form a 1D electrical contact along the edge of graphene with specific contact resistances as low as $\rho_c = 150$ Ωμm [15,16].

Inspired by the method of forming edge-contacts to graphene/hBN heterostructures, here we report a microfabrication strategy for the scalable microfabrication of electrical contacts to 2D materials based on the tri-layer resist system[17], that does not require encapsulation of the 2D material by hBN. Recently, we have used this method to nearly 3,000 contacts to graphene and demonstrated the largest, most accurate array of graphene quantum resistors[18], using epitaxial graphene on silicon carbide (epigraphene)[19]. This demanding application requires 100% yield, i.e. the total contribution of all contact resistances has to be well below 100 nano Ohms. To demonstrate the versatility of the edge-contact fabrication method we expand its use to another 2D material platform, the transition metal dichalcogenides (TMDCs) molybdenum disulfide ($MoS_2$). Our proof-of-concept tests on TMDCs show that our edge-contact fabrication process results in $MoS_2$-based field effect transistors (FET) with state-of-the-art performance: linear current-voltage (I-V) characteristic, ON/OFF ratio of $>10^6$ at room temperature, and current densities as large as 416 µA/µm at T=300 K (761 µA/µm at T=90 K) at large bias.

Figure 1a shows the principle of the edge-contact microfabrication method using the tri-layer resist system. The key to ensure scalable and reproducible contacts is to form a narrow/wide/narrow opening in the resist, which we achieve in a single e-beam exposure and single developing step by using a resist with higher sensitivity (compared to top and bottom layers) as the middle layer (see Methods and Supplementary S1). After resist development step, a dry etching step is used to remove the 2D material and expose its edges for a subsequent metallization step and metal lift off. For epigraphene, dry etching is achieved with oxygen plasma, while for $MoS_2$ the best results are obtained by performing a mild in-situ argon ion milling step before deposition of metals. The resist profile provides a clean lift-off when the metal deposition is by thermal evaporation and also by sputtering, giving additional flexibility when designing a microfabrication process.

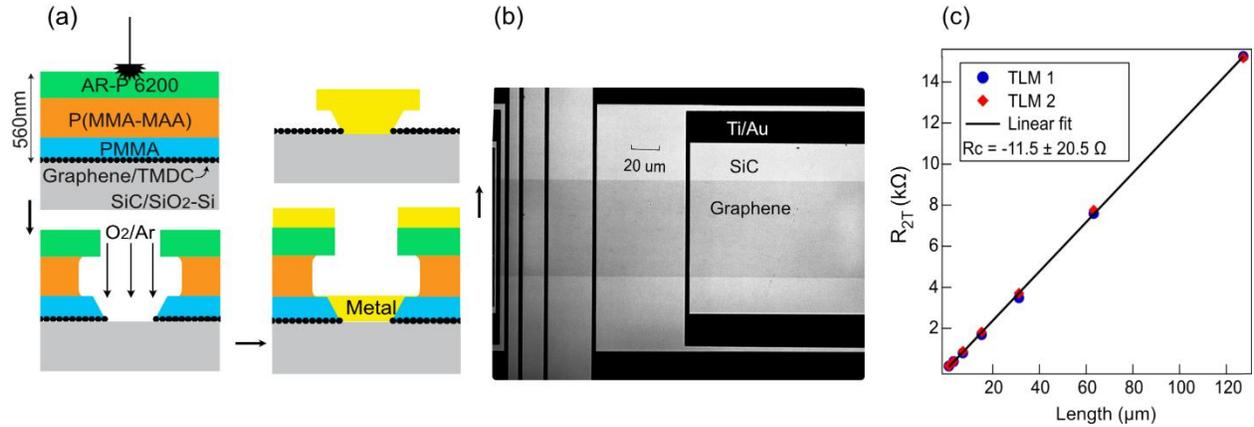

**Figure. 1.** Formation of scalable edge contacts to 2D materials. (a) Schematic representation of the tri-layer resist method for fabrication of edge contacts to 2D materials. (b) Optical image of a TLM device with edge-contacted graphene. (c) Example of TLM data for two devices with 64 um wide epigraphene channels. Note that the corresponding fit yields a negative (unphysical) $R_c$, close to zero value, with uncertainty $\pm 20.5\ \Omega$.

To illustrate the low contact resistance obtained using this fabrication method, we start describing the work performed on epigraphene, a wafer-scale material that allows us to fabricate large, ~mm-sized devices. For initial characterization we used the standard transfer length method (TLM) [20]. For this, graphene test structures of channel width $W$ are fitted with electrical contacts that are separated by a varying distance $L$ (Figure 1b). The measured two-terminal resistance $R_{2T}$ between adjacent contacts is:

$$R_{2T} = 2R_m + 2R_c + R_{channel} \quad (1)$$

which includes the (known) contribution of the metallic leads $R_m$, the metal-graphene interface $R_c$, and the graphene channel resistance, $R_{channel} = \rho\, L/W$, with $\rho$ the resistivity of graphene. A plot of the measured $R_{2T}$ versus the contact spacing will be in principle a straight line, where the resistivity of the material can be found from the slope, $dR_{2T}/dL = \rho/W$, and the contact resistance can be extracted from the intercept at $L = 0$, $R_{2T}(L = 0) = 2R_m + 2R_c$.

In total, we have performed TLM measurements on 12 epigraphene structures patterned on five epigraphene chips. Prior to measurements, samples were encapsulated to control and stabilize the doping of epigraphene[21]. The graphene channel width $W$ varied from 8-64 μm and the distances $L$ varied from 1-128 μm. Figure 1c shows an example of a TLM plot, $R_{2T}$ vs $L$, for the two largest epigraphene devices, $W = 64$ μm and total length of 270 μm. In this plot, each resistance datapoint is extracted from a linear fit to current-voltage curves (I-V) using a voltage bias up to $V = 1$ mV. The first observation is that slope of the two devices is remarkably similar, $dR_{2T}/dL = 119$ Ω/μm, giving a sheet resistance of 7616 Ω /□. As for the intercept, our linear fit yields a value of $R_c$ very close to zero with an uncertainty of ± 20.5 Ω. Quantitively, this indicates that the edge-contacts provides an electrical interface with very low contact resistance, and this is achieved for all the fabricated TLM structures with good reproducibility (See supplementary S2). The uncertainty in the TLM measurements is relatively low, not uncommon, and it may arise from e.g. local variations in the material doping, or the geometry of the fabricated devices.

To accurately quantify the low $R_c$ of our edge contacts, we turn to a Hall bar geometry and exploit the possibility offered by epigraphene to characterize the transparency of metal-graphene interface using a three-probe measurements $R_{3T}$ under quantum Hall effect conditions (QHE)[7,22]. In this three-probe measurement scheme (Figure 2a), the total measured resistance is:

$$R_{3T} = 2R_m + R_c + R_{channel}, \quad (2)$$

At low temperatures and in quantizing magnetic field[23], the graphene channel resistance vanishes, $R_{channel} = 0$, and a three-terminal measurement gives a direct measurement of the contact resistance. Figures 2b, c show the configuration and outcome respectively of the quantum Hall characterization of devices performed prior resistance of contact resistance, to ensure that the channel resistance $R_{channel} = Vxx/Ixx = 0$, and the transversal resistivity ($R_{xy} = Vxy/Ixx$) acquires the quantized value of $R_{xy} = h/2e^2 \approx 12.9$ kΩ, where $h$ is the Plank's constant, and $e$ is the elementary charge.

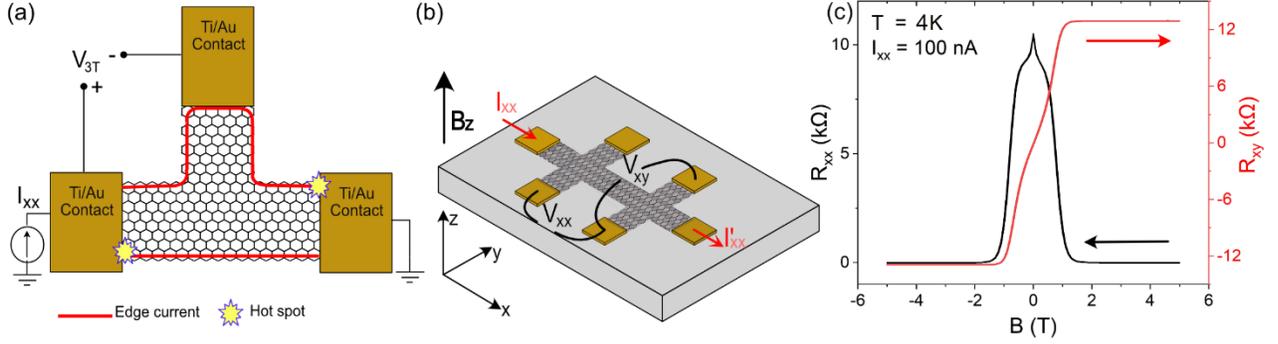

**Figure. 2.** Illustration of three and four terminal resistance measurements of epigraphene in quantum Hall regime. (a) Schematic of three terminal contact measurement on a Hall bar. In quantum Hall regime, the channel resistance vanishes due to the formation of edge states, and the measured three-terminal resistance, $R_{3T} = V_{3T}/I_{xx}$ includes the contact resistance $R_c$ and the (known) resistance of the leads $2R_m$. (b) Schematic of the measurements to extract the longitudinal resistance $R_{xx} = Vxx/Ixx$ and transverse resistance $R_{xy} = Vxy/Ixx$ measurement on a Hall bar. (c) Measured $R_{xx}$ (black) and $R_{xy}$ (red) in quantum Hall regime. In our devices, $R_{xx} = 0$ is obtained at B > 2T, and the QHE persists to temperatures up to T= 8 K and 100 nA.

Figure 3a shows that the typical phenomenology of edge contacts, i.e. the linear dependence of $R_c$ (Ω) on the inverse contact perimeter $1/l$, in fact extends to graphene in QHE[24,25]. These measurements were performed on 5 devices placed on the same chip, where the carrier density is tuned to n ≈ $10^{10}$-$10^{11}$ cm$^{-2}$. For comparison among devices placed on different chips (9 devices in 3 different epigraphene chips), we calculate and present the specific contact resistance $\rho_c = R_c \times l$, with $l$ the perimeter length. Figure 3b shows a histogram of $\rho_c$ obtained for nearly 70 Ti/Au edge contacts to epigraphene, with a mean value is $\overline{\rho_c} = 82$ Ωµm and a median value at $\widetilde{\rho_c}$ =67 Ωµm. These statistics, indicate that our tri-layer resist method can reproducibly produce contacts with low contact resistance. The spread in contact resistance values arises from the different doping of epigraphene in different chips. This is shown in Figure 3c, which displays the specific contact resistance data as a function of the carrier density $n$ measured in the devices from the low-field magnetoresistance data $n = 1/R_H$, with the Hall coefficient $R_H = dR_{xy}/dB$. The specific contact resistance follows the functional form for the Landauer quantum limit of resistance[5], $\rho_c^{min} = h/(2e^2 k_F) \approx 12906.403/\sqrt{n_{2D}}$ Ωµm. As the charge density is reduced below $1 \times 10^{11}$ cm$^{-2}$, the larger error bars in our measurements can be attributed due to larger effect of (local) charge inhomogeneity at the graphene-metal interface (charge puddles) [26],[27]. As a note, the $R_c$ measured in three-probe geometry in QHE arises from the region where current injection takes place, the so-called "hotspot", where all the dissipation and voltage drop takes place (see figure 3a) [28],[29]. From the $\propto l^{-1}$ and $\propto n^{-1/2}$ dependence of contact resistance, it appears that at the hotspot regions, electron transfer at the interface between metal and graphene in high magnetic field (i.e. in QHE) keeps similarities with electron injection into graphene at B=0. The main difference we observe around this region, is an effective carrier density about an order of magnitude higher than that extracted from the low field, $n = 1/R_H$.

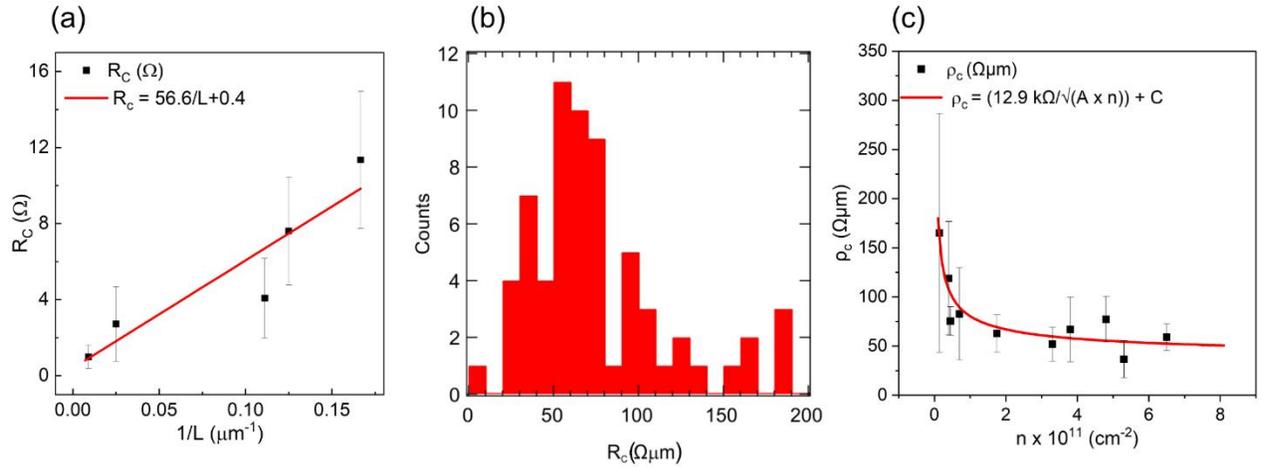

**Figure. 3.** Contact resistances $R_c$ of edge contacts to epigraphene measured in three-probe configuration in quantum hall regime. (a) $R_c$ (Ω) as a function of inverse perimeter length for 5 devices placed on the same chip. The linear fit yields a contact resistance $R_c(L) = 56.6/L + 0.4$ Ω. (b) Histogram of almost 70 edge contacts for 9 devices placed on 3 different epigraphene chips.

Contacts with higher $\rho_c$ >110 Ωμm are due to lowly doped graphene $n < 1 \times 10^{11}$ cm$^{-2}$. (c) $\rho_c$ as a function of carrier concentration. Solid line is fit to $\rho_c = h/(2e^2\sqrt{n_{eff}}) + C$, with $n_{eff} = A * n$. Here, $n_{eff}$ is an effective carrier density at the carrier injection points (hot spots). Our fits return the values A= 19.7 and $C = 34.5$ Ωμm. All error bars represent one standard deviation from the mean value.

Having demonstrated the low contact resistance reproducibly achieved for epigraphene, we now turn to the description of edge contact performance in semiconducting MoS₂. We observed that the uncertainties in TLM measurements on MoS₂ flakes resulted in much larger uncertainties compared to epigraphene devices, so we assessed the quality of the metal-TMDC interface in a field effect transistor (FET) geometry. In these devices, we evaluated key performance indicators: linearity of the current voltage characteristic, the transistor ON/OFF ratio, maximum current, mobility of the charge carriers in the FET.

Figure 4a shows a typical FET device prepared using a few-layer MoS₂ flake (about three layers, thickness 2.5 nm) as the channel material. Ti (10 nm)/Au (50 nm) have been evaporated to form 4 channels of length L = 1, 2, 3, and 4 μm, limited by the size of the flakes (2.5 μm² to 10 μm²).

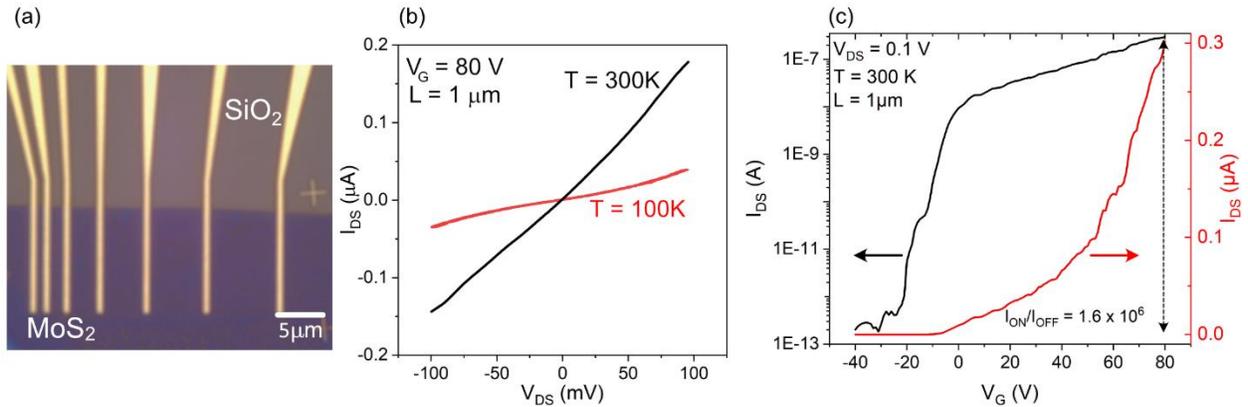

**Figure 4**. Edge contacts to molybdenum disulfide (MoS₂). a) Optical microscope image of the fabricated MoS₂ FET device. b) I-V measured in the ON state (gate voltage V$_G$ = 80 V) at room temperature and T=100 K. (c) Transfer characteristic of the MoS₂ device for $V_{DS} = 0.1$ V, T=300 K.

Figure 4b shows the I-V characteristic of the MoS$_2$ FET for 1μm channel length measured in the ON state ($V_G = 80$ V) for temperatures T = 100 K and 300 K. (See supplementary S3 for other channels lengths). The linearity of the I-V curves for all channel lengths indicate that edge-contacts are Ohmic. Figure 4c shows the transfer characteristic of MoS$_2$-FET at room temperature with an ON/OFF ratio is $I_{on}/I_{off} > 10^6$ (See supplementary S4 for measurements at high bias and low temperatures). These values match with the state-of-the-art TMDC field effect devices reported [8,30]. These measurements also allow to extract the field effect mobility, using $μ_{FE} = 1/qn_{2d}R_{sh} \approx L_{ch}g_m/WC_{ox}V_{ds}$, where $L_{ch}$, $W$, $C_{ox}$, $g_m$ are channel length, width, gate capacitance per area, and transconductance defined by $g_m = \partial I_d/\partial V_g$ at a constant $V_{DS}$. The calculated field effect mobilities reach 20 cm$^2$/V·s at 300 K. Altogether, the FET performance observed in our devices suggested the high-quality electrical interface achieved with our edge-contact method.

Finally, we present electron transmission microscopy (TEM) analysis of the edge contacts formed to epigraphene. Figure 5a shows an illustration of the edge contact formation between epigraphene and Ti/Au, as revealed by TEM. For this material, our process results in a contact to graphene such that the edge of the crystal and a small portion of its top surface (d < 20 nm) are directly contacted by the metal adhesion layer (Ti). Figure 5b shows an overview TEM image of the area around the graphene-metal contact. Figure 5c is a zoomed-in TEM image of a region ~ 30 nm away from the metallization area (yellow square in Fig. 5b), showing the presence of the two carbon layers of epigraphene, i.e. buffer layer in contact with SiC, and graphene as the topmost layer. The two carbon layers are atomically flat and continuously cover the whole SiC substrate, except the contact region. Regarding the details of the metal-graphene interface, we have used scanning TEM (STEM) - electron energy loss spectroscopy (EELS) to determine the boundary where graphene starts to be in contact with metal (Ti). Figure 5d is a false color map of the STEM-EELS chemical mapping at the contact edge, and it shows that graphene (green) is indeed contacted at the edge, but also part of its top surface is in contact with the titanium (red), over an extent d ~18nm. This chemical analysis establishes that the metal-graphene contact region occurs in the area approximately denoted by the dashed box in fig. 5b.

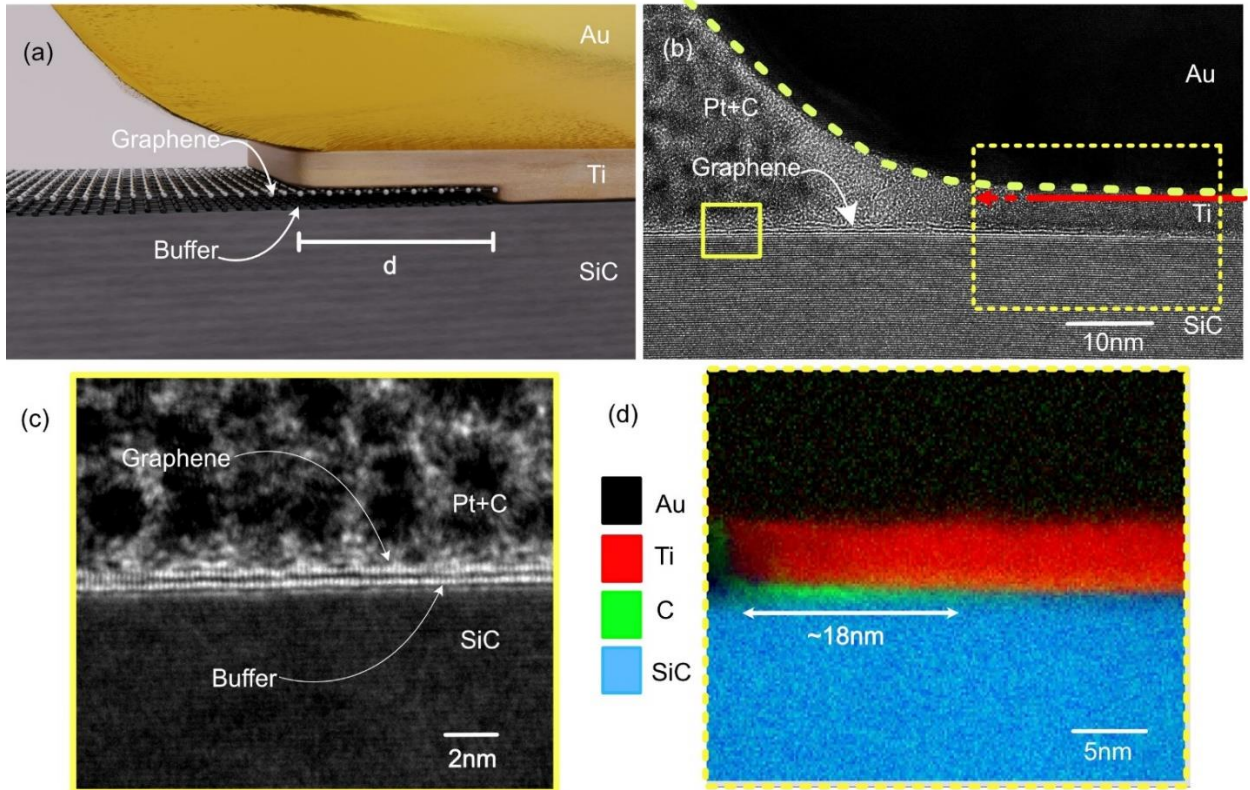

**Figure. 5.** TEM analysis of contacts to epigraphene. (a) Illustration of the graphene and buffer layer in contact with Ti/Au contact. (b) TEM image showing epigraphene-metal contact. The Pt+C layer is a protection layer deposited during TEM sample preparation. (c) TEM image of the area enclosed by the yellow rectangle in b, showing the presence of epitaxial graphene and buffer over SiC. (d) STEM-EELS chemical mapping of the interface between graphene and metal contacts. The chemical analysis reveals that epigraphene is contacted at the edge, and that Ti covers the graphene top surface over a distance d ~18 nm.

We foresee that the method to form scalable edge contacts for epigraphene and $MoS_2$ can be readily applicable to a wide variety of materials, including bulk and low dimensional materials.

Here demonstrated using e-beam lithography, the narrow/wide/narrow can also be achieved by optical lithography methods with some efforts and careful choice of resist layers. The possibility to achieve metal deposition by thermal evaporation and also by sputtering gives an additional advantage when designing electrical interfaces, which is key in exploring the electrical properties of emerging materials.

**METHODS**

**Materials:** $MoS_2$ flakes were exfoliated and transferred onto $Si-SiO_2$ substrates using the scotch tape technique and polydimethylsiloxane (PDMS) dry transfer. The $SiO_2$ is about 300 nm-thick. Typical thickness of thickness of prepared flake is 2 nm ± 0.5 nm. Epigraphene was grown on 4H–SiC chips (7x 7 mm2), which were encased in a graphite crucible and heated by RF heating to around 1850 C in an inert argon atmosphere of 1 bar.

**Devices:** For contact formation, we deposit a tri-layer of commercially available electron beam sensitive resists: 100 nm thick PMMA, 280 nm thick P(MMA-MAA) copolymer and 180 nm thick AR-P 6200.13 (see Figure 1a). After e-beam patterning, the top-most layer (AR-P 6200.13) is developed using o-xylene. For the bottom resist layers, we exploit the different sensitivities of PMMA and P(MMA-MAA) to the same developer, 93% IPA and 7% $H_2O$. This ensures the desired undercut that facilitates lift-off. After the development is completed, epitaxial graphene chips is etched by 1 minute of $O_2$-plasma at 50 W and 250 mT. For $MoS_2$ chips, in-situ Ar-plasma etching (1 minute) is performed before metal deposition. This step is achieved in a Lesker PVD225 deposition system equipped with an end-Hall linear ion source, operating at 170 V, 2 Amps and $2.2 \times 10^{-4}$ Torr. Metals deposition is performed at at pressure P = 5 x $10^{-7}$ mbar. A second lithography step and $O_2$-plasma are used for graphene to define the geometry of devices.

**Measurements:** Standard electrical characterization was performed in a liquid in a liquid 4He gas flow cryostat, which allowed for temperatures down to 2 K and magnetic fields up to 14 T. All reported values of charge carrier mobility and charge carrier concentration were extracted from four-probe Hall and quantum Hall measurements. The standard measurement setup used current biased samples at maximum of 100 nA DC (Keithley 6221 DC and AC current source, Agilent 34420 A nanovolt meter).

**Transmission electron microscopy:** TEM lamellas of the contacts were prepared using a FEI Versa focus ion beam – scanning electron microscope (FIB-SEM). Final polishing of the lamellas in FIB-SEM were performed with ion beam energy of 2 kV and beam current of ~ 20 pA to minimize beam effect. For TEM analysis, a JEOL monochromated ARM200F TEM was operated at 200 kV. The TEM is equipped with a CEOS GmbH probe corrector, a double silicon drift detector (SSD) for X-ray energy dispersive spectroscopy (XEDS), and a Gatan GIF Continuum for electron energy loss spectroscopy (EELS). STEM-EELS and STEM-XEDS spectrum imaging measurements are used for elemental mapping.

**Corresponding Author**

Author to whom correspondence should be addressed: samuel.lara@chalmers.se

**Author Contributions**

The manuscript was written through contributions of all authors. All authors have given approval to the final version of the manuscript.

**Funding Sources**

This work was jointly supported by the Swedish Foundation for Strategic Research (SSF) (nos. GMT14-0077, RMA15-0024), Chalmers Excellence Initiative Nano, 2D TECH VINNOVA

Behavior of the contacts of quantum Hall effect devices at high currents *J. Appl. Phys.* **96** 404–10

[30] Nourbakhsh A, Zubair A, Sajjad R N, Amir T K G, Chen W, Fang S, Ling X, Kong J, Dresselhaus M S, Kaxiras E, Berggren K K, Antoniadis D and Palacios T 2016 MoS2 Field-Effect Transistor with Sub-10 nm Channel Length *Nano Lett.* **16** 7798–806